\documentclass{article}
\usepackage{arxiv}
\usepackage[utf8]{inputenc} 
\usepackage[T1]{fontenc}    
\usepackage{hyperref}       
\usepackage{url}            
\usepackage{booktabs}       
\usepackage{amsfonts}       
\usepackage{nicefrac}       
\usepackage{microtype}      
\usepackage{lipsum}
\usepackage{graphicx}
\usepackage{amsmath}
\usepackage[round,sort&compress,comma,numbers]{natbib}
\usepackage{color}
\newcommand{\ed}[1]{\textcolor{black}{#1}}

\title{DeepMoD: Deep learning for Model Discovery in noisy data}

\author{
  {Gert-Jan Both} \\
  CRI Research \\
  Universite Paris Descartes \\
  Paris, France \\
  \And
  Subham Choudhury \\
  CRI Research \\
  Universite Paris Descartes \\
  Paris, France \\
  \And
  Pierre Sens \\
  Institut Curie, CNRS UMR168 \\
  PSL Research University \\
  Paris, France \\
  \And
  Remy Kusters\thanks{Corresponding Author.} \\
  CRI Research \\
  Universite Paris Descartes \\
  Paris, France \\
  \texttt{remy.kusters@cri-paris.org}\\
}

\begin{document}
\maketitle

\begin{abstract}

\ed{We introduce DeepMoD, a Deep learning based Model Discovery algorithm. DeepMoD discovers the partial differential equation underlying a spatio-temporal data set using sparse regression on a library of possible functions and their derivatives. A neural network is used as function approximator and its output is used to constructs the function library, allowing to perform the sparse regression \emph{within} the neural network.} This construction makes it extremely robust to noise, applicable to small data sets, and, contrary to other deep learning methods, does not require a training set. We benchmark our approach on several physical problems such as the Burgers', Korteweg-de Vries and Keller-Segel equations, and find that it requires as few as $\mathcal{O}(10^2)$ samples and works at noise levels up to $75\%$. Motivated by these results, we apply DeepMoD directly on noisy experimental time-series data from a gel electrophoresis experiment and find that it discovers the advection-diffusion equation describing this system.

\end{abstract}


\section{Introduction}
Recently, efforts have been made to combine data-driven science with bottom up physical modelling in a new field known as "theory-guided data science" \cite{karpatne2017}. Integrating first-principle models with data science has already proven successful in material science \cite{wagner2016}, earth science \cite{karpatne2017lake, reichstein2019deep, kong2018, de2017deep} and fluid mechanics \cite{maulik2017, miyanawala2017}. This approach has proven useful to infer coefficients of known PDEs from artificial data, the so called Physics-Informed Neural Networks (PINNs) \cite{raissi2017I, raissi2017II,raissi2018deep, raissi2019physics}, and even to directly discover physical models from artificial data, i.e., PDE-NET \cite{long2017pde, long2018pde}, PDE-Stride \cite{maddu2019} and PDE-Find \cite{rudy2017, rudy2018}.

\begin{figure*}
    \centering
    \includegraphics[width=1\linewidth]{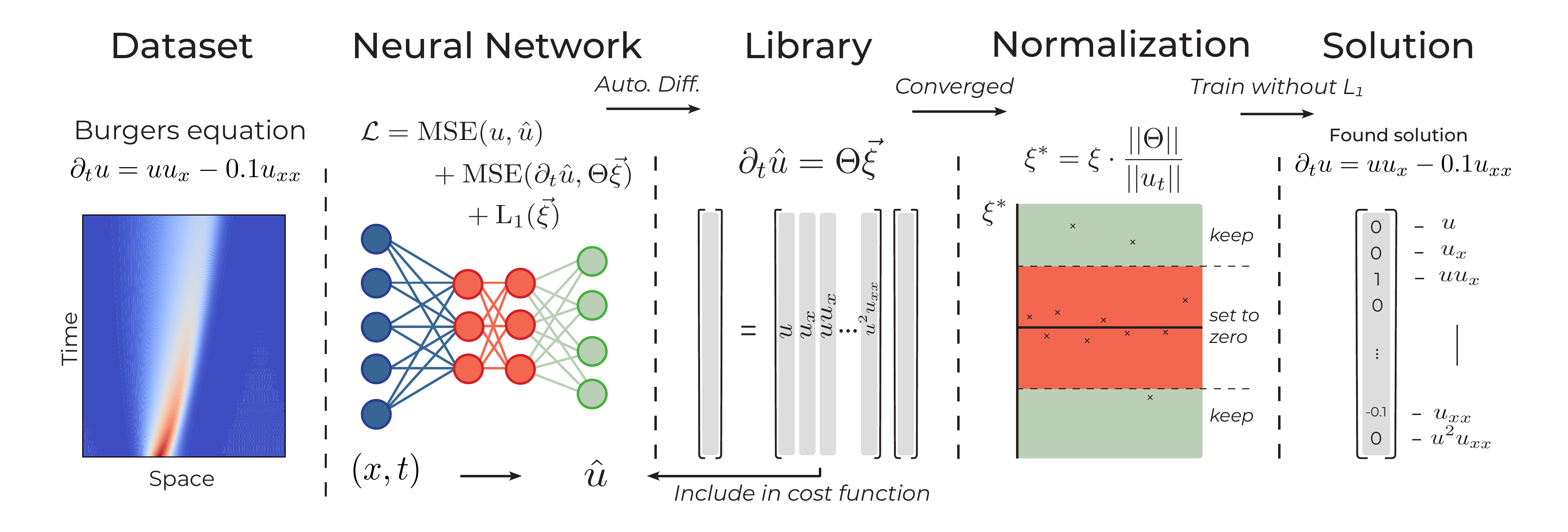}
    \caption{Work-flow of the algorithm: the neural network takes the coordinates of the problem as input and approximates the noisy dataset as output. From this output, the library of candidate terms is constructed and subsequently included in the regression part of cost function of the network, which consists of the MSE-loss, a regression loss and an $L_1$ regularization. Once the neural network has converged or reached the maximal numbers of allowed iterations, the elements of $\vec{\xi}$ are normalized and thresholded. The resulting sparsity pattern is then used to train the network one final time without $L_1$ penalty to find the unbiased coefficients.}
    \label{fig:1}
\end{figure*}

The problem of data-driven model discovery of PDEs has been approached from several different directions. While information theory provides a rigorous basis for model selection, it becomes computationally infeasible to compare the information criteria of a vast amount of candidate models \cite{mangan2016}. Alternatively, an approach to discover a PDE from a spatio-temporal data-set is to use sparse regression model selection schemes such as PDE-FIND as proposed by \cite{rudy2017, rudy2018}. In this approach, the PDE underlying a dataset $u(\{x,t\})$ is discovered by writing the model discovery task as a regression problem,
\begin{equation}
    \partial_t u = \Theta \xi,
    \label{eq:intro}
\end{equation}
where $\Theta$ is a matrix containing a library of polynomial and spatial derivative functions (e.g. $u, u_x, u u_x$). Here model discovery turns into finding a sparse representation of the coefficient vector $\xi$. Rudy et al. \cite{rudy2017} introduce the regression algorithm TrainSTridge to solve this task on artificial data such as the Burgers' equation in hydrodynamics, the Schr\"{o}dinger equation in quantum mechanics and the Kuramoto-Sivashinsky equation in chaos theory \cite{rudy2017}. Although very promising, this method is sensitive to noise and requires a large number of samples. This can largely be traced back to errors in the numerical differentiation and hence inaccurate derivatives in the library $\Theta$. These traits essentially render the method unfeasible on noisy experimental data. To overcome this, we propose to use automatic differentiation within the neural network to accurate calculate the derivatives in the library function. Indeed, Rudy et al. \cite{rudy2017} recognize this possibility to improve the performance of exisiting model discovery methods such as SINDY \cite{mangan2016} and PDE-find \cite{rudy2017}. A first approach would be to use a neural network to learn the mapping of the data, i.e. $\{x,t\} \to u$, and then employ automatic differentiation to accurately calculate the derivatives of $u$ with respect to $x$ and $t$, which can then be used to construct $\Theta$. Unfortunately, this implementation is susceptible to overfitting noisy data, which significantly decreases the accuracy of the library. 

The novelty of our work is that we circumvent this ubiquitous issue by implementing Eq. \ref{eq:intro} \emph{within the cost function of the neural network}. Consequently, training the network not only adjusts the weights and biases of the network, but also adjusts the components of the sparse vector $\xi$, corresponding to Eq. \ref{eq:intro}. An $L_1$ term on $\xi$ is added to the cost function to ensure its sparsity. Training the neural network yields the underlying PDE and denoises the data set. We show that this approach outperforms state-of-the-art methods of model selection \cite{rudy2017,raissi2019physics} by applying it on artificial data sets of the Burgers', Korteweg-de Vries (KdV), 2D advection diffusion and the Keller-Segel equations (See SI for a comparison). Finally we demonstrate that DeepMoD can discover the PDE underlying an electrophoresis experiments and discover the 2D advection diffusion equation. This shows that this deep learning based selection algorithm can consistently discover the second order advection diffusion equation directly from a simple time-series of images of a diffusing dye.

\section*{Methods}

Our goal is to develop a fully-automated procedure which discovers the partial differential equation (PDE) underlying a measured data set. Given a data set $u(\{x,t\})$, we can write this problem as,

\begin{equation}
   \partial_t u(x,t) = \mathcal{F}\left(u, u_x, u u_x, u_{xx}, ... \right),
   \label{eq:u_t}
\end{equation}

where we seek the function $\mathcal{F}$. To find $\mathcal{F}$, we generate a large set of possible models by considering all permutations of a library of candidate terms. The choice of the library depends on the problem at hand but generally consists of polynomial basis functions and their corresponding spatial derivatives. For example, in the one dimensional examples we present in this paper, the library consists of all polynomials in $u$ up to second order, the derivatives of $u$ with respect to the coordinates (e.g. $\partial_x u$) up to third order and any possible combinations of these two sets (e.g. $u^2u_{xx}$), totalling just 12 terms. However, one can construct more than 4000 unique models from this limited set of functions, rendering an information theory approach computationally unfeasible \cite{mangan2016}. 
 
We circumvent this problem by utilizing a sparse regression approach, in which the model discovery problem is rewritten as
 
\begin{equation}
    \partial_t u = \Theta \xi,
    \label{eq:eq1}
\end{equation}

where $\partial_t u$ is a column vector of size $N$ containing the time derivative of each sample and $\Theta$ contains all $M$ possible terms for each of the samples, so that we can write it as

\begin{equation}
\small{
    \Theta = \begin{bmatrix}
    1 & u(\{x,t\}_0) & u_x(\{x,t\}_0) &  \dots & u^2 u_{xx}(\{x,t\}_0)   \\
    1 & u(\{x,t\}_1) & u_x(\{x,t\}_1) &  \dots & u^2 u_{xx}(\{x,t\}_1)  \\
     \vdots    &   \vdots  & \vdots & \ddots & \vdots\\
         1 & u(\{x,t\}_N) & u_x(\{x,t\}_N) &  \dots & u^2 u_{xx}(\{x,t\}_N) \\
\end{bmatrix}}
\end{equation}

Since $\Theta$ contains significantly more terms than required, most coefficients in $\xi$ will be zero and hence we are looking for a sparse solution of the vector $\xi$. \ed{Note that since we directly discover the governing PDE rather than a closed form of the solution, the differential equation that we discover is independent of the precise boundary conditions of the problem at hand.} In the next section, we discuss how this regression task is solved using Lasso, a sparsity promoting regression method, \emph{within a neural network}.

\subsection*{Lasso in neural network} 

\ed{In order to solve the regression task of the previous section we need to contruct the function library, $\Theta$. Here we employ a densely-connected feed-forward neural network which takes the spatial and temporal coordinates of the problem, i.e. $\{x, t\}$ as input, and outputs $\hat{u}$, an approximation of  $u$ at $\{x,t\}$ \cite{raissi2017I,raissi2017II}. In other words, the neural network approximates the function $u(x,t)$ and employs this approximation to construct the library function, $\Theta$. Using feed-forward neural networks as function approximators has three major advantages, i) they naturally accommodate non-linear constraints, without the need to linearize any operators, ii) they do not require any time-stepping scheme and iii) they allow the use of automatic differentiation to accurately differentiate the output of the neural network with respect to the input coordinates.}

The neural network we consider here is trained by optimizing the cost function,

\begin{equation}
\mathcal{L} = \mathcal{L}_{MSE} + \mathcal{L}_{Reg} + \mathcal{L}_{L_1}.
\label{eq:L}
\end{equation}

\begin{figure}[h!]
    \centering
    \includegraphics[width=0.98\linewidth]{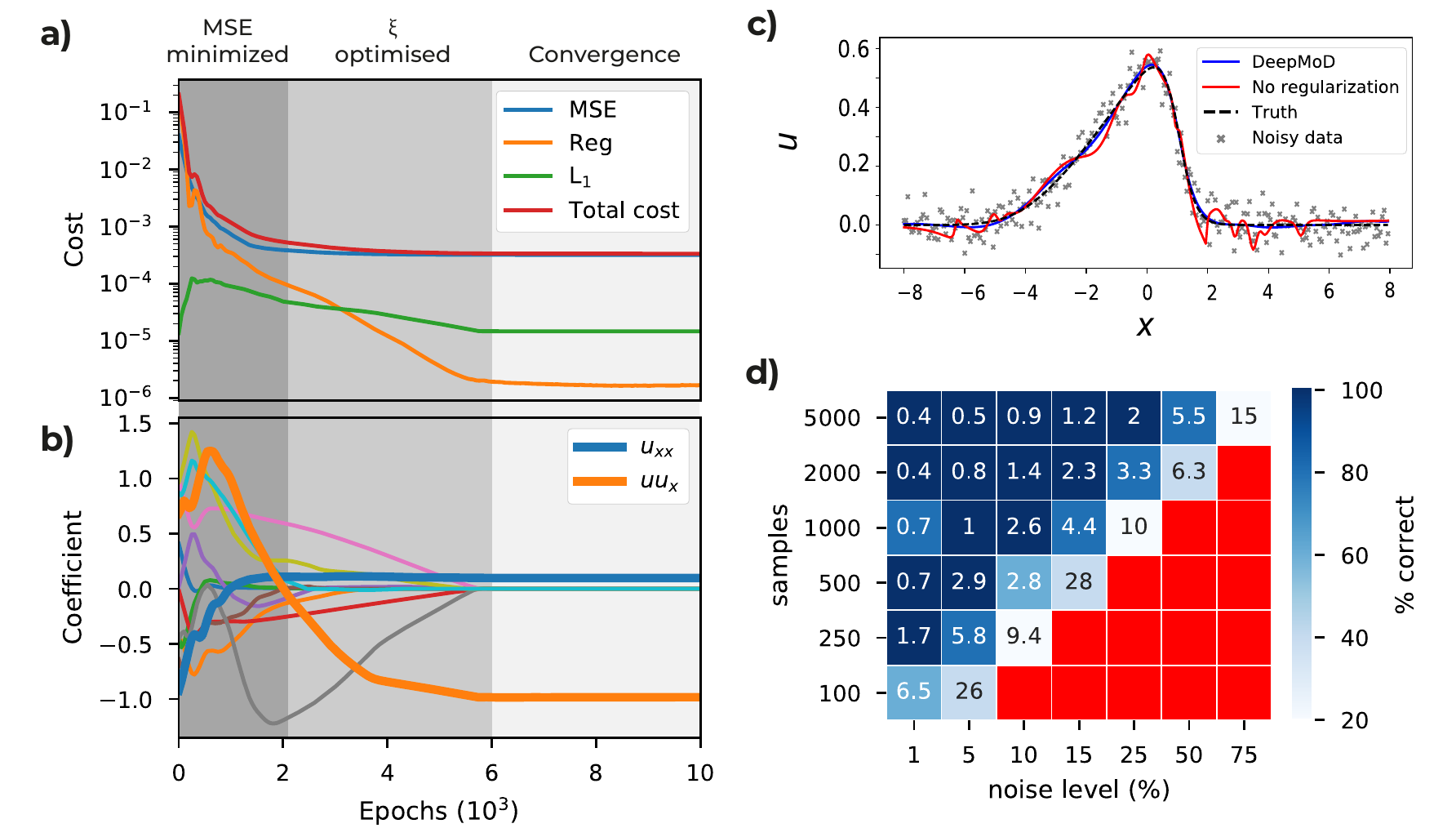}
    \caption{\textbf{(a)} Cost functions and \textbf{(b)}  coefficient values, $\xi$, as function of the number of epochs for the Burgers' dataset consisting of 2000 points and $10\%$ white noise. Initially, the neural network optimizes the MSE and only \emph{after} the MSE is converged the coefficient vector is optimized by the network. \textbf{(c)} Velocity field $u$ for $t = 5$ obtained after training \emph{with} (no overfitting) and \emph{without} (overfitting) the regression regularization $\mathcal{L}_{Reg}$. \textbf{(d)} The values in the grid indicate the accuracy of the algorithm tested on the Burgers' equation, defined as the mean relative error over the coefficients, as function of the sample size of the data set and level of noise. The coloring represents the fraction of correct runs (Red indicates that in none of the five iterations the correct PDE is discovered).}
    \label{fig:2}
\end{figure}

Here, $\mathcal{L}_{MSE}$, is the mean squared error (MSE) of the output of the neural network $\hat{u}$ with respect to the dataset $u(\{x,t\})$,
\begin{equation}
\mathcal{L}_{MSE} = \frac{1}{N}\sum_{i=1}^{N}|u(\{x, t\}_i) - \hat{u}_i|^2.
\label{eq:mse}
\end{equation}
The last two terms of Eq. \ref{eq:L} correspond to the Lasso regularization: $\mathcal{L}_{Reg}$ performs regression to find the coefficient vector $\xi$ and $\mathcal{L}_{L_1}$ is an $L_1$ regularizer on $\xi$. In order to implement the regression problem (Eq. \ref{eq:eq1}) within the neural network, we introduce the regression based cost function,
\begin{equation}
\mathcal{L}_{Reg} = \frac{1}{N}\sum_{i=1}^{N}| \Theta_{ij}\xi_j- \partial_t \hat{u}_i|^2. 
\label{eq:costPI}
\end{equation}
Note that the coefficient vector $\xi$ is updated alongside the weights and biases of the neural network, while the terms in $\Theta$ are computed from the output of the neural network (i.e. $\hat{u}$). Automatic differentiation is used to calculate all the spatial and temporal derivatives in $\Theta$, returning machine-precision derivatives. This approach is considerably more accurate than any form of numerical differentiation. Moreover, $\mathcal{L}_{Reg}$ acts as a regularizer on $\hat{u}$, preventing overfitting of the noisy data set, even though our library contains a large amount of terms (See Results and Fig. \ref{fig:2}c).

Finally, an $L_1$ regularization on the vector $\xi$ is added to ensure its sparsity,

\begin{equation}
\mathcal{L}_{L_1} = \lambda \sum_{i=2}^{M}|\xi^i|.
\end{equation}

Here $\lambda$ is a constant setting the strength of the regularization (further discussed in the SI). 

The total cost of the neural network is then minimized using the Adam optimizer. The combination of the MSE term and the regression term in the cost function constrain the network in such a way that it converges to the right solution. To determine if the network has converged, we introduce a convergence criterion. As we show in Fig. \ref{fig:2}(a,b), the MSE converges before $\xi$ does, so that our criterion is based on the convergence of $\xi$:

\begin{equation}
    \text{max}\left(\frac{\partial \mathcal{L}}{\partial \xi_i}\frac{||\partial_t u||}{||\Theta_i||}\right) < \rm{tol}.
\end{equation}
 
This criterion states that the maximum value of the gradient of the loss function with respect to the coefficients must be smaller than a given tolerance. Note here that we have scaled the gradients as we discuss in the next paragraph. Since it is not guaranteed the network will reach this tolerance, we train the network until the convergence criterion is satisfied, or for a maximum amount of iterations. 

\subsection*{Normalization and thresholding} 
When the neural network has finished training, we obtain the sparse vector $\xi$. Despite the $L_1$ regularization, most terms will be non-zero and hence we need to threshold the small coefficients to obtain the true sparse representation. Since each term has different dimensions, Eq. \ref{eq:u_t} is rendered dimensionless,

\begin{equation}
    \partial_t u \to \frac{\partial_t u}{||\partial_t u ||}, \Theta \to \frac{\Theta}{||\Theta||} ,\xi \to \xi \frac{||\Theta||}{||\partial_t u||},
\end{equation}

where $||\Theta||$ is the norm of each column of $\Theta$ and $||\partial_t u||$ the norm of the time-derivative vector. As a result of this transformation, components of $\xi$ will typically be $\mathcal{O}(1)$.

\begin{figure}
    \centering
    \includegraphics[width=0.8\textwidth]{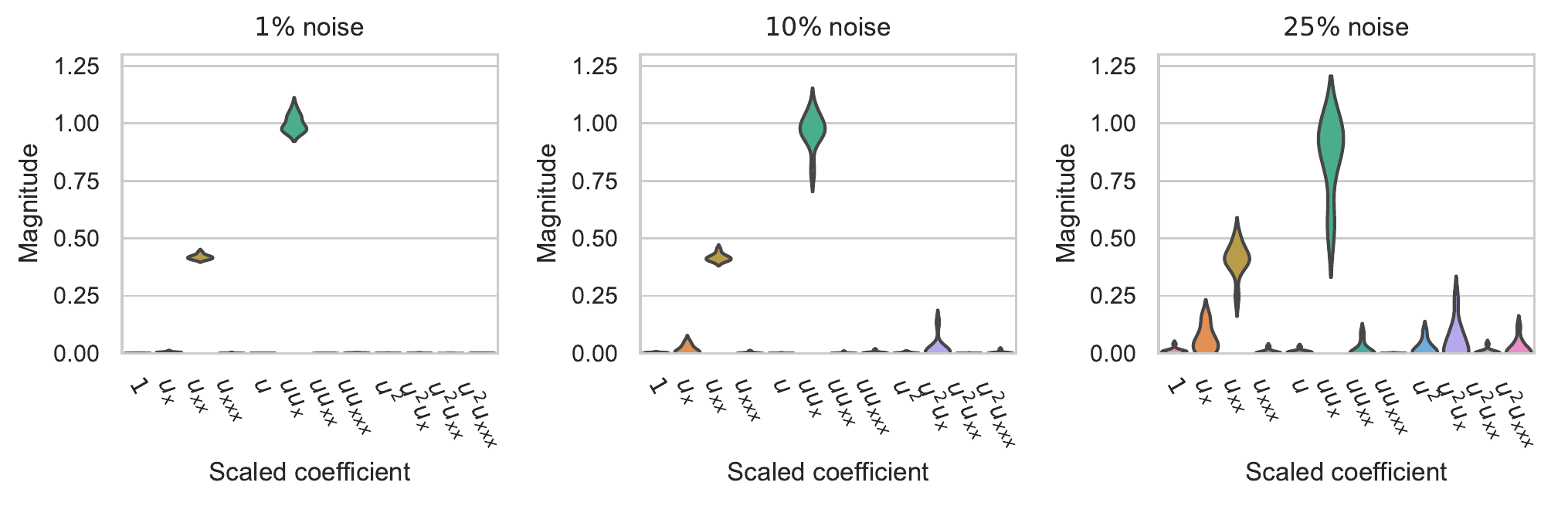}
    \caption{Scaled coefficients before thresholding for the Burgers equation with $1\%$ noise (\textbf{a}), $10\%$ noise (\textbf{b}) and $25\%$ (\textbf{c}). This indicates that there is a considerable difference between the terms that do and do not feature in the PDE.}
    \label{fig:threshold}
\end{figure}

\ed{Thresholding prunes components with a negligible impact on the data set by setting all their values to zero. Figure \ref{fig:threshold} shows the distribution of the scaled coefficients before thresholding for the $1\%$, $10\%$ and $25\%$ noise runs with sample sizes of $2000$ and $5000$ (corresponding to the data in figure 2D). Figure \ref{fig:threshold} shows that the required terms (i.e. $u_{xx}$ and $uu_{x}$) considerably stand out relative to the others. We show that up to moderate noise levels $<20\%$ (Fig. a and b) the exact value of the threshold does not significantly impacts which terms are eventually pruned. The difference between the largest non-featuring and smallest featuring terms is typically up to an order of magnitude. For very high noise levels $> 20\%$ levels (See Fig. \ref{fig:threshold} c) the accuracy of the pruning is much more sensitive to the exact value and more advanced sparsity algorithms could be required to obtain more robust model selection (See Discussion). 
}

We then train the network one final time without $L_1$ penalty and with the regression term only containing the terms selected in the first cycle, to find an unbiased estimate of the coefficients of the underlying PDE.
\section*{Results}

We test the performance of DeepMoD on a set of case studies: the Burgers' equation with and without shock, the Korteweg - de Vries equation, the 2D advection-diffusion equation and the Keller-Segel model for chemotaxis. These examples show the ability of DeepMoD to handle (1) non-linear equations, (2) solutions containing a shock wave,  (3) coupled PDEs and finally (4) higher dimensional and experimental data. 

\subsection*{Non-linear PDEs}

We apply DeepMoD to recover various non-linear and higher order differential equations. As examples we consider Burgers' equation (in the SI we the Korteweg-de Vries equation, which contains a third-order derivative). The Burgers' equation occurs in various areas of gas dynamics, fluid mechanics and applied mathematics and is evoked as a prime example to benchmark model discovery \cite{rudy2017, long2018pde} and coefficient inference algorithms \cite{raissi2017I, raissi2017II, raissi2019physics}, as it contains a non-linear term as well as second order spatial derivative,
\begin{equation}
    \partial_t u = - u u_x + \nu u_{xx}. 
    \label{eq:Burgers}
\end{equation}
Here $\nu$ is the viscosity of the fluid and $u$ its velocity field. We use the dataset produced by Rudy et al. \cite{rudy2017}, where $\nu=0.1$. The numerical simulations for the synthetic data were performed on a dense grid for numerical accuracy. DeepMoD requires significantly less datapoints than this grid and we hence construct a smaller dataset for DeepMoD by randomly sampling the results through space and time. From now on, we will refer to randomly sampling from this dense grid simply as sampling. Also note that this shows that our method does not require the data to be regularly spaced or stationary in time. For the data in Fig. \ref{fig:2} we add $10\%$ white noise and sampled 2000 points for DeepMoD to be trained on. 

\begin{figure}[h!]
\centering
\includegraphics[width=0.5 \linewidth]{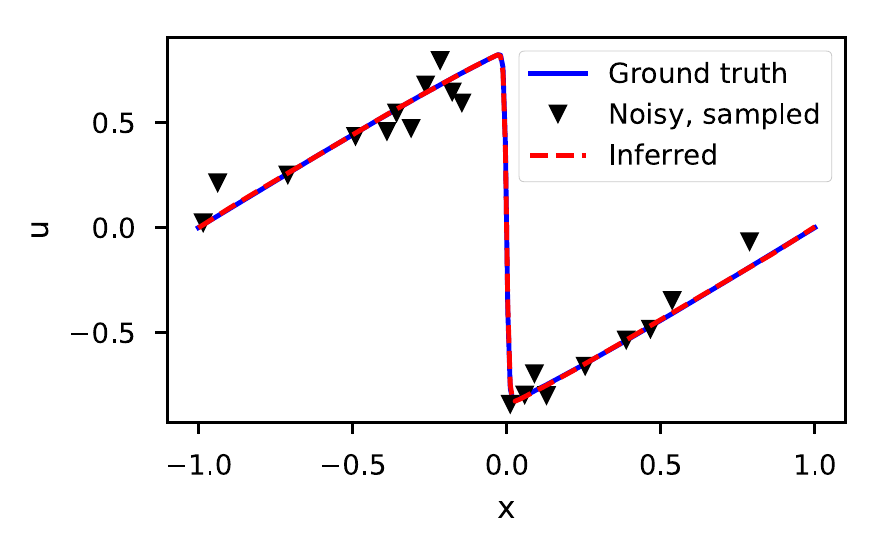}
\caption{Ground truth, Noisy and Inferred data at $t=0.8$ for the Burgers' equation with a shock wave (10$\%$ noise and 2000 sample points).}
\label{fig:3}
\end{figure}

We train the neural network using an Adam optimizer (see SI for details) and plot the different contributions of the cost function as a function of the training epoch in Fig. \ref{fig:2}a and we show the value of each component of $\xi$ as a function of the training epoch in Fig. \ref{fig:2}b. Note that for this example, after approximately 2000 epochs, the MSE is converged, while at the same time we observe the components of $\xi$ only start to converge after this point. We can thus identify three 'regimes': in the initial regime (0 - 2000 epochs), the MSE is trained. Since the output of the neural network is far from the real solution, so is $\Theta$, and the regression task cannot converge (See first 2000 epochs in Fig. \ref{fig:2}b). After the MSE has converged, $\hat{u}$ is sufficiently accurate to perform the regression task and $\xi$ starts to converge. After this second regime (2000 - 6000 epochs), all components of the cost converged ($>$6000 epochs) and we can determine the solution. From this, we obtain a reconstructed solution (see Fig. \ref{fig:2}b) and at the same time recover the underlying PDE, with coefficients as little as $1 \%$ error in the obtained coefficients. We show the impact of including the regression term in the cost function in Fig.\ref{fig:2}c, where the obtained solution of DeepMoD is compared with a neural network, solely trained on the MSE, to reconstruct the data. Including the regression term regularizes the network and prevents overfitting, despite the many terms in the library. We conclude that it is the inclusion of the regression in the neural network which makes DeepMoD robust to noisy data and prevents overfitting.

Next, we characterize the robustness of DeepMoD in Fig. \ref{fig:2}d, where we run DeepMod for five times (differently sampled data set) for a range of sample sizes and noise levels. The color in Fig. \ref{fig:2}d shows how many of the five runs return the correct equation and the value in the grid displays the mean error over all correct runs. Observe that at vanishing noise levels, we recover Eq. \ref{eq:Burgers} with as little as 100 data-points, while for 5000 data points we recover the PDE with noise levels of up to $75\%$. Between the domain where we recover the correct equation for all five runs and the domain where we do not recover a single correct equation, we observe an intermediate domain where only a fraction of the runs return the correct equation, indicating the importance of sampling (See SI 2 for further discussion). 

\ed{To benchmark DeepMoD, we can directly compared the performance of our algorithm with respect to two state of the art methods, (i): PDE-Find by Rudy et al. \cite{rudy2017} and (ii) PDE-Stride by Maddu et al. \cite{maddu2019}. We considered an identical Burgers’ data set and for $10^5$ data points, approach (i) recovers the correct equation for up to $1\%$ Gaussian noise  \cite{rudy2017} while method (ii) discovers the correct equation up to $5\%$ noise \cite{maddu2019}. Compared to the results in Fig. \ref{fig:2}d we note that even for two order of magnitude fewer samples points, $10^3$ w.r.t. $10^5$, DeepMoD recovers the correct equation up to noise levels $> 50 \%$ Gaussian noise. DeepMoD allows up to two orders of magnitude higher noise-levels and smaller sample sizes with respect to state-of-the-art model discovery algorithms. The reason DeepMoD is considerable more robust w.r.t. noise and sample size is two-fold: (i) numerical differentiation or denoising requires a relatively fine sampling grid to accurately aproximate the derivatives present in the library function. (ii) Since the functions in the library of the neural network are calculated with respect to the inferred solution, our approach is considerably less sensitive to elevated noise levels (See Fig. \ref{fig:2}c). We show in the SI that DeepMod has similar performance for the KdV equation, which contains a third order spatial derivative.}  

\subsection*{Shock wave solutions}

If the viscosity is too low, the Burgers' equation develops a discontinuity called a shock (See Fig. \ref{fig:3}). Shocks are numerically hard to handle due to divergences in the numerical derivatives. Since DeepMoD uses automatic differentiation we circumvent this issue. We adapt the data from Raissi et al \cite{raissi2017II}, which has $\nu=0.01/\pi$, sampling 2000 points and adding 10 $\%$ white noise (See Fig. \ref{fig:3}). We recover ground truth solution of the Burgers' equation as well as the corresponding PDE, 
\begin{equation}
    \partial_t u = - 0.99 u u_x + 0.0035 u_{xx},
\end{equation}
with a relative error of 5$\%$ on the coefficients. In Fig. \ref{fig:3} we show the inferred solution for $t=0.8$.

\begin{figure}
    \centering
    \includegraphics[width=0.6 \linewidth]{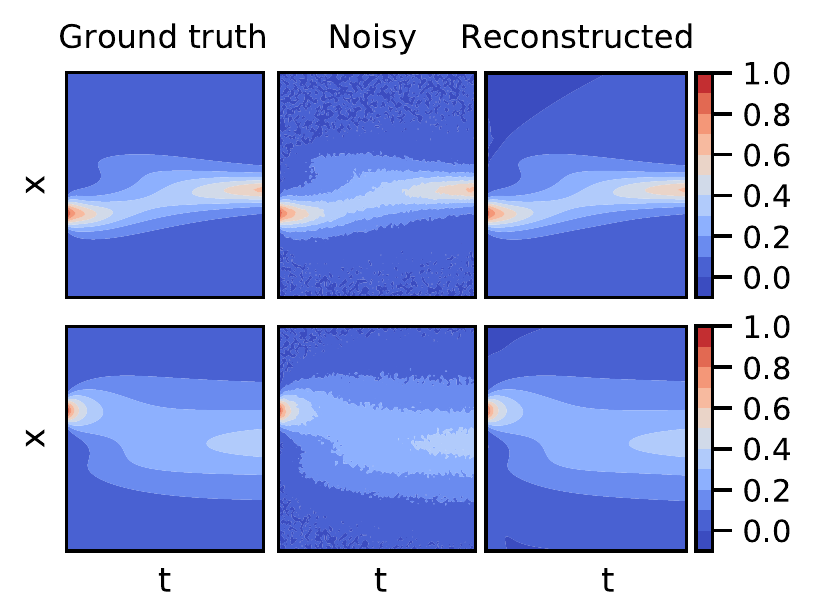}
    \caption{Ground truth, noisy and reconstructed solutions for the density of cells, $u$ (top row) and the density of secreted chemicals $w$ (bottom row) in the Keller Segel model for 5 $\%$ white noise and 10000 samples.}
    \label{fig:KellerSegel}
\end{figure}

\subsection*{Coupled differential equations}
Next, we apply DeepMod to a set of coupled PDE's in the form of the Keller-Segel (KS) equations, a classical model for chemotaxis \cite{keller1970, chavanis2010}. Chemotactic attraction is one of the leading mechanisms that accounts for the morphogenesis and self-organization of biological systems. The KS model describes the evolution of the density of cells $u$ and the secreted chemical $w$,
\begin{equation}
\begin{split}
    \partial_t u = \nabla \cdot \left( D_u \nabla u - \chi u \nabla w \right) \\
    \partial_t w = D_w \Delta w - k w + h u.
\end{split}
\end{equation}
Here the first equation represents the drift-diffusion equation with a diffusion coefficient of the cells, $D_u$ and a chemotactic sensitivity $\chi$, which is a measure for the strength of their sensitivity to the gradient of the secreted chemical $w$. The second equation represents the reaction diffusion equation of the secreted chemical $w$, produced by the cells at a rate $h$ and degraded with a rate $k$. For a 1D system, we sample 10000 points of $u$ and $w$ for parameter values of $D_u = 0.5$, $D_v = 0.5$, $\chi = 10.0$, $k = 0.05$ and $h = 0.1$ and add 5 $\%$ white noise. We choose a library consisting of all spatial derivatives (including cross terms) as well as first order polynomial terms, totalling 36 terms. For these conditions we recover the correct set of PDEs, 
\begin{equation}
\begin{split}
    \partial_t u = 0.50 u_{xx} - 9.99 u  w_{xx}  -  10.02  u_x  w_x\\
    \partial_t w = 0.48 w_{x x} - 0.049 w + 0.098 u,
\end{split}
\end{equation}
 
as well as the reconstructed fields for $u$ and $w$ (See Fig. \ref{fig:KellerSegel}). Note that even the coupled term, $u_x  w_x$ , which becomes vanishingly small over most of the domain, is correctly identified by the algorithm, even in the presence of considerable noise levels. 

\subsection*{Experimental data}

To showcase the robustness of DeepMoD on high-dimensional and experimental input data, we consider a 2D advection diffusion process described by,
\begin{equation}
    \partial_t u =  -\nabla\cdot\left(-D\nabla u + \vec{v} \cdot u \right),
    \label{Eq:AdvectionDiffusion}
\end{equation}
where $\vec{v}$ is the velocity vector describing the advection and $D$ is the diffusion coefficient. In the SI we apply \emph{DeepMod} on a simulated data-set of Eq. \ref{Eq:AdvectionDiffusion}, with as initial condition, a 2D Gaussian with $D = 0.5$ and $\vec{v} = (0.25,0.5)$. For as little as 5000 randomly sampled points we recover the correct form of the PDE as well as the vector $\vec{v}$ for noise levels up to $\approx 25 \%$. In the absence of noise the correct equation is recovered with as little as 200 sample points through space and time (See SI 2). This number is surprisingly small considering this is an 2D equation. 

\begin{figure}[h!]
\centering
\includegraphics[width=0.6 \linewidth]{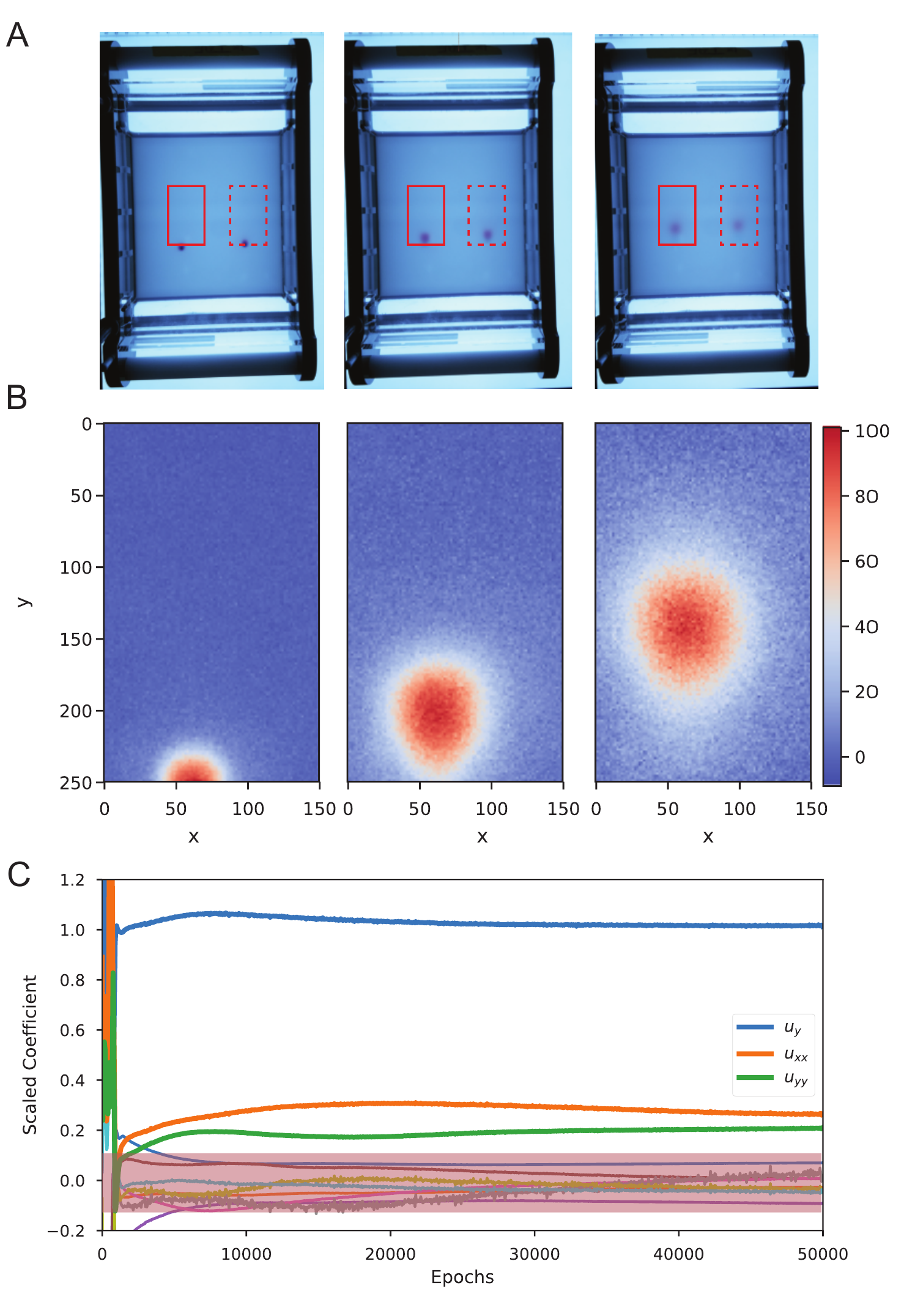}
\caption{\textbf{(a)} Time-serie images of the electrophoresis essay. The two red boxes indicate the analysed region. \textbf{(b)} Region indicated in the solid red box of (a) showing the density of the dye at three different time-frames (in pixels). \textbf{(c)} Scaled coefficients values of all the candidate for a single run. The pink region indicates the terms with scaled coefficient $| \xi | < 0.1$.}
\label{fig:6}
\end{figure}

Finally, we apply DeepMoD on a time-series of images from an electrophoresis experiment, tracking the advection-diffusion of a charged purple loading dye under the influence of a spatially uniform electric field (See SI for further details). In Fig. \ref{fig:6}a we show time-lapse images of the experimental setup where we measure the time-evolution of two initial localised purple dyes. Fig. \ref{fig:6}b shows the resultant 2D density field for three separate time-frames (in arbitrary units), corresponding to the red square in Fig. \ref{fig:6}a by substracting the reference image (no dye present). The dye displays a diffusive and advective motion with constant velocity $v$, which is related to the strength of the applied electric field.

We apply DeepMoD on 5000 sampled data-points sampled through space and time and consistently recover the advection term $u_y$ as well as the two diffusive components ($u_{xx}$ and $u_{yy}$). In Fig. \ref{fig:6}c we show the scaled coefficients as function of the number of epochs. After thresholding the scaled coefficients ($|\xi| < 0.1$), we obtain for thre unscaled coefficients, the resultant advection diffusion equation,
\begin{equation}
    0.31 u_y + 0.011 u_{xx} + 0.009 u_{yy} = 0.  
\end{equation}

Analysing the second diffusing dye (dashed box in Fig. \ref{fig:6}) result in nearly identical values for the drift velocity, $v \approx 0.3$, and the diffusion coefficients, $D \approx 0.01$ (See SI) indicating the robustness of the obtained value of $D$ and $v$. In contrast to the artificial data presented in previous paragraphs, some higher-order non-linear terms, in particular $u u_{yy}$ and $u u_{xx}$ remain small, yet non-zero. This suggests that an automatic threshold strategy may not guarantee the desired sparse solution. Fixing a threshold of the scaled coefficients ($|\xi| < 0.1$ in this particular case) or other thresholding strategies such as coefficient cluster detection would be better suited for this task.

\section*{Discussion}
In this paper we presented \emph{DeepMoD}, a novel model-discovery algorithm, utilising neural networks to discover the underlying PDE of a spatio-temporal dataset. We demonstrate the algorithm on 5 artificially obtained case studies: Burgers' (with/without shock), Korteweg-de Vries, advection diffusion and Keller-Segel equations as well as on an experimental data-set of the advection diffusion equation. In contrast to many of the state of the art model discovery algorithms DeepMoD is very robust with respect to elevated noise levels and resilient to small data set sizes, demonstrating an automated model selection task directly from an experimental obtained time-series measurement. DeepMoD allows higher dimensional input/output as well as coupled PDEs as demonstrated with the 2D advection diffusion and Keller-Segel equation.

Through the use of automatic differentiation, combined with a regression-based regularization, the approximation of the spatio-temporal derivatives in noisy data is strongly enhanced. DeepMoD combines two previously established ideas, (i) a regression-based approach to model discovery (pioneered by e.g. Rudy et al. \cite{rudy2017, rudy2018}) and (ii) the ability of neural networks to infer system parameters in the context of Physics Informed Neural Networks (Raissi et al. \cite{raissi2017I, raissi2017II, raissi2019physics}. We show that combining both approaches strongly improves the model discovery task at hand and results in an increased robustness with respect to noise-levels and sample size for model discovery tasks. This approach, for the first time, allows model selections on highly noisy and hence low spatial/temporal resolution experimental data, which to date is one of the prime challenges of this field. DeepMoD also allows to infer the various type of diffusive, chemo-tactic equations, from single particle tracking (SPT) data by following a similar approach as \cite{rudy2017}, which will advance existing approaches to infer potentially anomalous diffusive processes from SPT data \cite{el2015, Granik2019}. 

The success of this approach however strongly relies on (i) the completeness of the library functions in $\Theta$ and (ii) a threshold of small yet non-zero terms. (i) If the underlying functions are not present in the library, DeepMoD will not return the correct underlying equation. This problem however can be identified via a cross-validation procedure on a set on a spatial/temporal domain that is not present in the training data. If correct, the resultant equation should perform well outside the spatio-temporal domain of training. \ed{Conversely, since we use neural networks as function approximator, the function library, $\Theta$, can be tailored to the problem at hand and thus contain non-linear functions of the network's output, e.g. $\sin u$, $1/(1+u)$. This can be used to model the spatio-temporal evolution of e.g. genetic activation networks \cite{crombach2014}.} We have empirically found that including these extensive libraries does not result in over-fitting the sparse coefficient vector of the data, even though the optimisation contains more degrees of freedom. \ed{(ii) While the threshold criterion based on the standard deviation of the coefficient vector provides consistent results throughout the artificial data-sets, this approach understandably fails when either the data is very noisy or when experimental artefacts introduce non-zero contributions of higher order terms. While we have shown in Fig. \ref{fig:threshold} that up to moderate noise levels the exact value of the threshold does not impact the results significantly, for very high noise levels more advanced sparsity selections algorithms like PDE-find \cite{rudy2017} or coefficient clustering schemes would be more appropriate.}

Besides the model selection capabilities, DeepMoD demonstrates its usefulness to denoise data and allows accurately approximating derivatives from noisy data, a notoriously difficult task to solve with classical interpolation and finite difference schemes. Employing this "function library based" regulation of neural network architecture may boost the enhancement of e.g. super resolution images through physics informed regularisation \cite{yang2010image, von2019}.

\section*{Acknowledgements} Thanks to the Bettencourt Schueller Foundation long term partnership, this work was partly supported by CRI Research Fellowship to Remy Kusters. We acknowledge the support of NVidia through their academic GPU grant program. We thank Jonathan Grizou for his valuable input on the manuscript, Ariel Lindner and Pascal Hersen for suggesting the gel electrophoresis experiments and Thea Chrysostomou for helping with the experiments.

G.B. and R.K. developed the code and tested it on all examples. S.C. helped testing the algorithm on the last example. P.S. and R.K. supervised the project. G.B. and R.K. wrote the paper.

Code and examples available at https://github.com/PhIMaL/DeePyMoD.

\bibliographystyle{unsrtnat}  
\bibliography{references}  

\end{document}


\maketitle

In this Supplementary Information we discuss DeepMoD and the synthetic datasets on which we have benchmarked the algorithm. To recapitulate, the core idea behind DeepMoD is to find a sparse representation of the underlying PDE by generating a \textit{library}, $\Theta$, composed of polynomial basis functions with their corresponding spatial derivatives. This essentially reduces the model selection task to  finding a sparse vector, $\xi$, corresponding to, 
\begin{equation}
\partial_t u = \Theta \xi,
\label{eq:reg}
\end{equation}
which best fits the data. DeepMoD aims to recover a sparse vector, $\xi$, using a neural network and by implementing the regression scheme within the network. The algorithm can be divided into three parts: First, a feed-forward neural network approximates the mapping from the input data $\{x,t\}$ to the output $u(\{x,t\})$. The derivatives of $u(\{x,t\})$ are determined from this mapping using automatic differentiation and are used to construct a library containing all possible terms in the PDE. The coefficient vector $\xi$ is obtained by performing Lasso directly within the neural network. After the total loss of the neural network has converged we threshold the resulting weight vector and obtain the sparse vector $\xi$. The final sparse weight vector only contains non-zero terms corresponding to active terms, i.e. the terms that feature in the PDE, with their corresponding value. In the following sections we discuss the various parts of the algorithm in detail. The full code as well as the data-sets are available on Github: https://github.com/PhIMaL/DeePyMoD.

\subsection*{Cost function}

A fully connected feed-forward network is used to map the input coordinates of the problem $\{x , ... t\}$ to a scalar or vector output $u(\{x , ... t\})$. We use the tanh activation function as it is infinitely differentiable; the chosen activation function has to be differentiable at least as many times as the highest order derivative. For example, a ReLu activation would henceforth not work for libraries containing second-order (and higher) derivatives. The neural network learns the mapping by minimizing the total cost function, which is comprised of the mean squared error, $\mathcal{L}_{MSE} $, a regression based cost function, $\mathcal{L}_{Reg}$ and a sparsity promoting cost function, $\mathcal{L}_{L_1}$,

\begin{equation}
    \label{eq:Cost_MSE_PI}
    \mathcal{L} = \mathcal{L}_{MSE} + \mathcal{L}_{reg} + \mathcal{L}_{L_1} =\frac{1}{N}\sum_{i=1}^N |u(\{x, t\}_i)-\hat{u}_i|^2 + \frac{1}{N}\sum_{i=1}^N |f(\{x, t\}_i)|^2 + \lambda \sum_{i=2}^M |\xi_i|\cdot \frac{||\Theta_i||}{||\partial_t\hat{u}||}
\end{equation}

The cost function related to the mapping $\{x , ... t\}$ to a scalar or vector output $u(\{x , ... t\})$ is written as,
\begin{equation}
\mathcal{L}_{mse} = \frac{1}{N}\sum_{i=1}^{N}|u(\{x, t\}_i) - \hat{u}_i|^2,
\label{eq:LSI}
\end{equation}
where $N$ is the number of samples. Automatic differentiation is then used to find the derivatives of the outputs with respect to the inputs, e.g.  $\partial_t u, u_x, u_{xx} ...$. These derivatives are then combined with polynomial basis functions to yield the function library $\Theta$. For a 1D+1 input $\{x,t\}$ and single output $\hat{u}$, we select a library constructing all possible combinations of spatial derivatives $\partial^n_x u $ up to order $n$ and polynomials $u^m$ up to order $m$, which, including a constant term, gives a library of $(n+1)\times(m+1) = M$ terms. We implement the regression problem by first rewriting it as $f = \partial_t u - \mathcal{F}(...)$ and include it in the cost function by minimizing,
\begin{equation}
\mathcal{L}_{reg} = \frac{1}{N}\sum_{i=1}^{N}|f(\{t,x\}_i)|^2.
\label{eq:costPISI}
\end{equation}

\begin{figure}
    \centering
    \includegraphics[width=0.8 \linewidth]{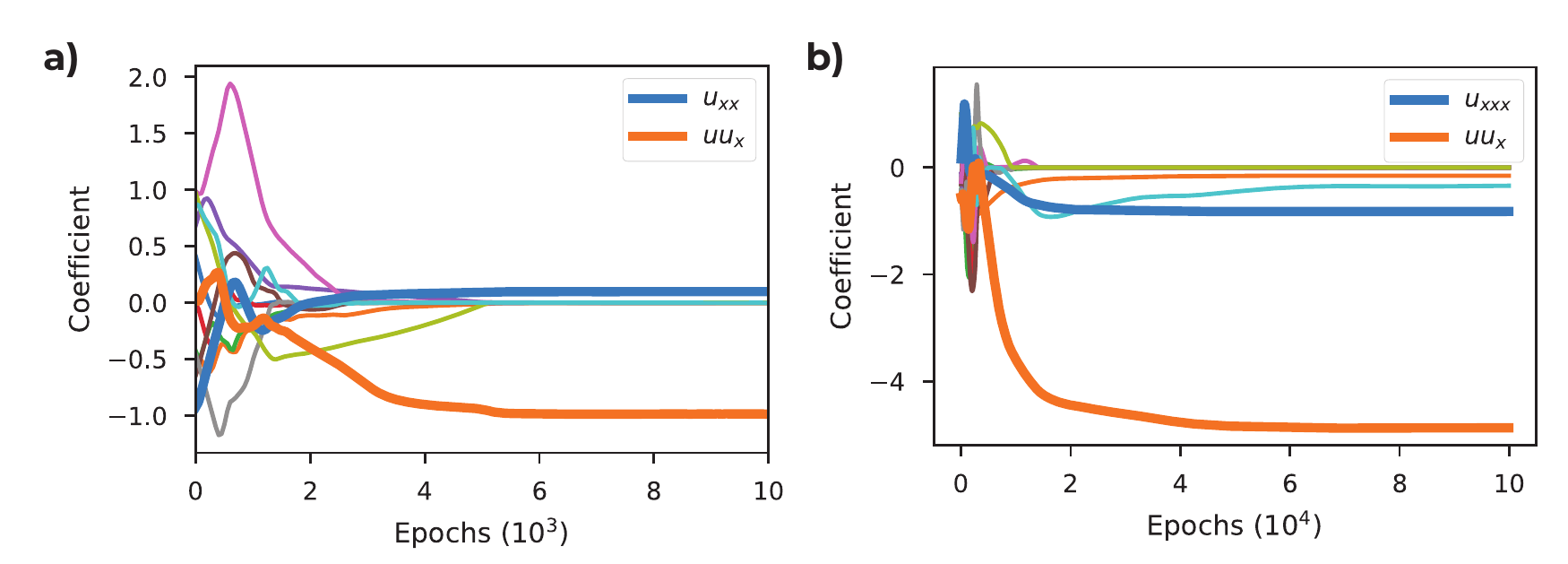}
    \caption{Evolution of unscaled components of $\xi$ for a library containing fourth order derivatives and third order polynomial. \textbf{(a)} shows the Burgers' equation, \textbf{(b)} the KdV, both for a 2000 point dataset with $10\%$ white noise.}
    \label{fig:biglibrary}
\end{figure}

To showcase that the algorithm is not sensitive to the number of terms in the library, we plot the evolution of the different terms of $\xi$ for a Burgers' and a Korteweg-de Vries (KdV) data set of 2000 data points and 10 $\%$ of white noise in Fig. \ref{fig:biglibrary}.

Here we increased the amount of library terms up to fourth order derivatives and third order polynomials, and thus a total of 20 terms, compared to 12 in the example in the main text. We still obtain the correct equation with identical accuracy of the obtained coefficients but including higher order derivatives increases the computational time significantly.

Finally, we add a sparsity-promoting component in the form of an L$_1$ regularization on $\xi^*$:
\begin{equation}
\mathcal{L}_{L_1} = \lambda \sum_{i=2}^{M}|\xi_i^*|.
\end{equation}

\begin{figure}
    \centering
    \includegraphics[width=0.8 \linewidth]{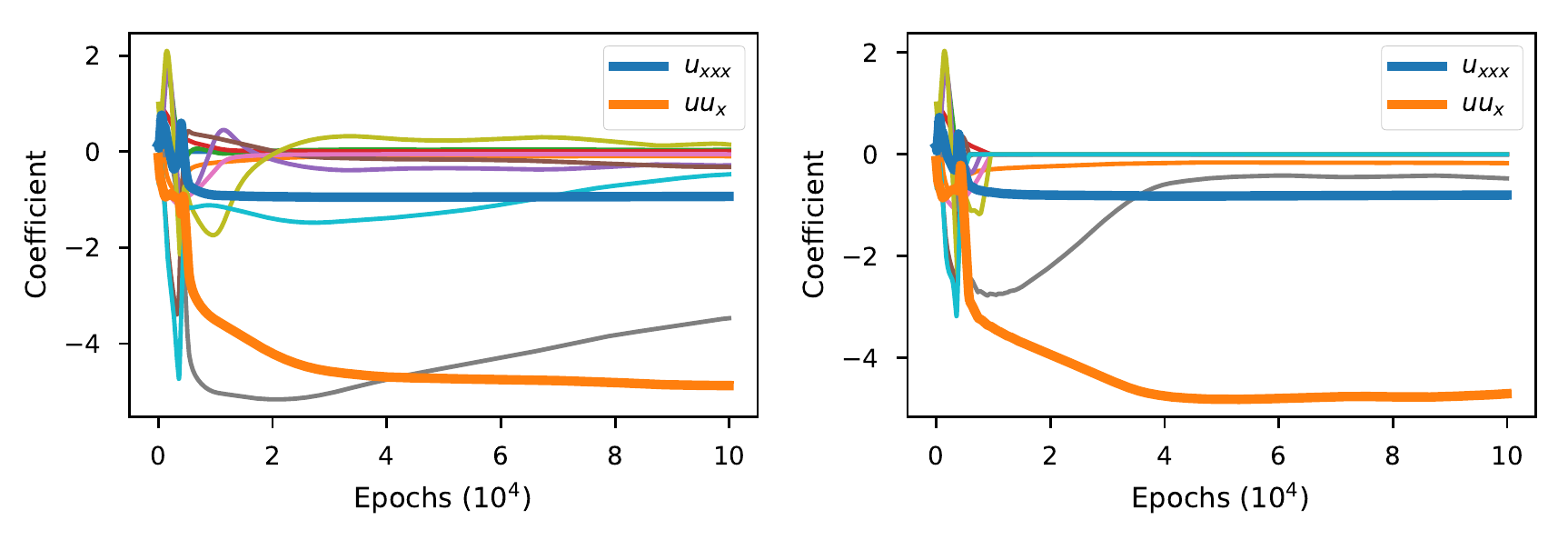}
    \caption{Evolution of unscaled components of $\xi$ for a KdV dataset containing 2000 points with $10\%$ white noise. \textbf{(a)} shows without L$_1$ regularization (i.e. $\lambda=0$), \textbf{(b)} shows the evolution with $\lambda=10^{-5}$.
    }
    \label{fig:l1}
\end{figure}

Here, $\xi_i^*$ represents the $i$-th element of the scaled coefficient vector (we discuss the scaling in the next paragraph). Note that the regularization is not applied to the constant term, hence the summation starts from $i=2$. The parameter $\lambda$ in front of this cost (See Eq. \ref{eq:Cost_MSE_PI}) is set to $\lambda = 10^{-5}$ for all the presented examples. To showcase the importance of including the L$_1$ regularization however, we show in Fig. \ref{fig:l1} the evolution of $\xi$ as function of the epoch for the KdV equation using a dataset of 2000 points and $10\%$ white noise. \textbf{(a)} Shows this without L$_1$ regularization and \textbf{(b)} including an L$_1$ term with $\lambda=10^{-5}$. Essentially it increases the convergence rate of the network: whereas the network with L$_1$ regularization has converged after roughly $60.000$ epochs, the one without is not converged after $100.000$ epochs.

\subsection*{Training the Neural network and the role of noise on the MSE}

To train the neural network, we minimize the total cost function using the Adam optimizer with slightly adjusted parameters for faster convergence: we use a increased learning rate of 0.002 (except of the Keller Segel where we used the default value: 0.001) and set $\beta_1$ to 0.99 (default value: 0.9). We also explored the L-BFGS-B optimizer and found that while it was faster than Adam, it sometimes suffered from convergence issues. 

\begin{figure}
    \centering
    \includegraphics[width=0.8 \linewidth]{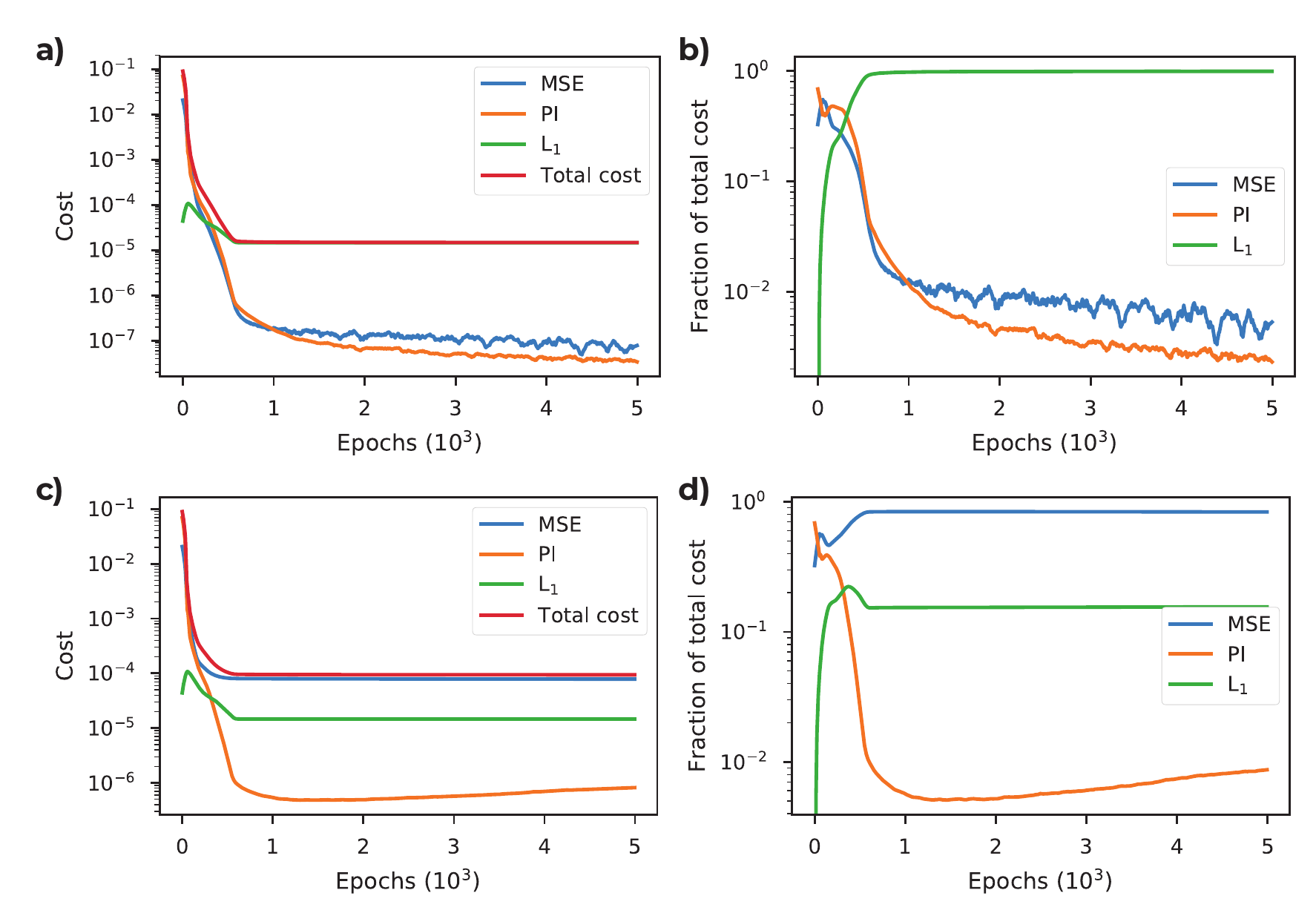}
    \caption{\textbf{a)} Absolute cost of each component of the cost function for a noiseless Burgers' dataset of 2000 datapoints.
    \textbf{b)} Relative cost of the Burgers' dataset of panel (a).
    \textbf{c)} Absolute cost of each component of the cost function for a noisy (5$\%$ white noise) Burgers' dataset of 2000 datapoints.
    \textbf{d)} Relative cost of the Burgers' dataset of panel (c). Note that these figures were smoothed using a first order Savitzky-Golay filter with a window length of 15.}
    \label{fig:cost}
\end{figure}

To gain more insight into the minimization of the neural network, we trace the different components of the cost function (i.e. the MSE, regression and L$_1$ part) as a function of the epoch for the Burgers dataset in Fig. \ref{fig:cost} for a data set without noise (a) and a dataset with  $5\%$ noise (2000 randomly sampled data points). Panels (a) and (c) contain the absolute cost whereas panels (b) and (d) plot the corresponding relative cost (the neural network consisted of five hidden layers of twenty neurons each and $\lambda = 10^{-5}$).

We find that for both datasets, the network converges after approximately 8000 epochs. For the noiseless case, we observe in panel (b) that the L$_1$ component makes up the largest fraction of the cost. In contrast, for the noisy case, the total cost is dominated by the MSE. For the noiseless case both MSE and the regression term should in principle approach zero, leaving the L$_1$ as the main component of the total cost. Given that the L$_1$ regularization acts on the scaled coefficients, which are on the order $\mathcal{O}(1)$ and that only few components should be significant, the total cost should be of the order of $\lambda$: $\mathcal{O}(10^{-5})$, which we indeed observe. In the case of the noisy data set however, the regression prevents overfitting and denoises the data. This results in a residual cost of the MSE cost (see panel (d)). In other words, the MSE cost is a measure of the difference between the true underlying and measured dynamics due to the noise. With this measure we can directly estimate the noise level from the MSE by writing the data $u(x,t)$ as the sum of the ground truth $u_0$ and additive Gaussian noise $\sigma \mathcal{N}(0,1)$, 

\begin{equation}
    MSE = \langle (\hat{u} - u_0 - \sigma \mathcal{N}(0,1))^2\rangle.
    \label{eq:MSE}
\end{equation}

If the network learns the \textit{exact} solution, $ \hat{u} \to u_0$, Eq. \ref{eq:MSE} becomes

\begin{equation}
    \label{eq:minMSE}
    MSE = \sigma^2 \langle \mathcal{N}^2(0,1)\rangle = \sigma^2.
\end{equation}

For panel (c), we obtain a value of $\approx 10^{-4}$, which indeed corresponds to Fig. \ref{fig:cost}(c). If a network converged at a MSE lower than this value, it would be a sign of overfitting; Eq. \ref{eq:minMSE} is the minimal MSE. We empirically tested the robustness against overfitting by making our network significantly larger (10 layers of 100 neurons) and did not observe overfitting. 

To determine if the network converged, we implement a convergence criterion. As stated in the main text, the MSE converges before the coefficients and we thus determine the convergence based on $\xi$:

\begin{equation}
    \text{max}\left(\frac{\partial \mathcal{L}}{\partial w_i}\frac{||\partial_t u||}{||\Theta_i||}\right) < \rm{tol}.
\end{equation}

We calculate the derivative of the loss function with respect to $\xi$ and normalize it to fairly compare the gradients. We then require the maximum value of this vector to be smaller than some tolerance, as all components have to be converged. Typically, we choose $\text{tol} = 10^{-6}$. Note however that while convergence is typically reached within $\pm 25000$ epochs it can sometimes take considerably longer for all the coefficients to properly converge, especially when the function library is very large (See e.g. the example in Fig. \ref{fig:biglibrary} for the KdV equations with an extended library). 

\subsection*{Scaling and thresholding}

After the network has converged we obtain the vector $\xi$. Despite the $L_1$ regularization, many terms are close to but non-zero. We thus need to threshold $\xi$ in order to obtain the true sparse vector. Since each component has different units and possibly different absolute scales, we normalize the components before thresholding. We calculate the norm of each column in $\Theta$ and the norm of the time derivative and scale $\Theta$ and $\xi$ accordingly: 

\begin{equation}
    \begin{split}
    &\Theta \to \Theta^{*}\cdot||\Theta||^{-1}, \\
    &\xi \to \xi^{*}\frac{||\Theta||}{||\partial_t u||},\\
    & \partial_t u \to \partial_t u^{*} ||\partial_t u||^{-1}
    \end{split}
\end{equation}

where $||\Theta||$ is a row vector containing the norms of each column in $\Theta$ and $\xi^{*}$ and $\Theta^{*}$ are the normalized weights vector and library. By normalizing accordingly, the regression problem will only consist of unit-length vectors, which allows for an easier interpretation of $\xi^*$; typically, it will be $\mathcal{O}(10^{-1} - 10^{0})$. The normalization also allows DeepMoD to select terms with different orders of magnitude. 

Next, we threshold this scaled vector. Since most components are close to zero, we interpret the thresholding operation as finding outliers from the median. We hence set any component smaller than the standard deviation $\sigma$ to zero:

\begin{equation}
    \xi_i=
    \begin{cases}
    \xi_i \text{ if } \xi^*_i > \text{median}(\xi^*) + \sigma(\xi^*)\\
    0 \text{ if } (\text{median}(\xi^*) - \sigma(\xi^*)) < \xi^*_i < (\text{median}(\xi^*) + \sigma(\xi^*))\\
    \xi_i \text{ if } \xi^*_i < \text{median}(\xi^*) - \sigma(\xi^*).\\
    \end{cases}
    \label{Eq:threshold}
\end{equation}

Although this criterion is somewhat arbitrary, in all the benchmarks we presented here it performs well. To showcase the importance of the normalization in the tresholding, we plot the normalized and unnormalized values of $\xi$, obtained from the Burgers equation with shock after $10^5$ epochs. For this dataset, the non-linear term and the diffusion term differ two orders of magnitude ($1$ versus $0.01/\pi$). Observe that due to the small value of the diffusion term it cannot be distinguished from the non-active terms. By normalizing the coefficients however, both components are significantly larger in magnitude relative to all other coefficients.

After the thresholding we have determined the sparsity pattern, i.e. which terms are present in the PDE. Due to our use of $L_1$ regularization our estimates of the coefficients are not unbiased however. We thus run the network again with only the components of the sparsity pattern and without $L_1$, similar to a Physics Informed Neural Network. 
\subsection*{Computational cost}
The computational cost of DeepMoD strongly depends on e.g. the size of the data set, the order of the derivatives, the hardware (CPU or GPU) and optimisation options. A typical run needs 5 - 10 minutes; on a machine with a CPU (1950X AMD Threadripper) with 16GB RAM took 3 minutes on a data set of 1000 points and 10 minutes on a data set of 5000 points for a third-order library. The increase in computational cost due to the increase in data set size can be circumvented by using a GPU, as long as the whole data set and model can be loaded into its memory. The bulk of the computational costs come from calculating the library, which is re-evaluated every single iteration. Higher order derivatives require subsequent backwards passes through the neural network, which is computationally expensive. Nonetheless, improved optimiser settings, smarter initialisation and potentially batching could further reduce the computational time.

\section*{Synthetic data}

DeepMoD was tested on five synthetic case studies: the Burgers' equation with and without shock formation, the Korteweg-de Vries equation (KdV), the 2D advection diffusion equation and the Keller-Segel equations. To allow a direct comparison with existing algorithms, we used the data-set provided in Rudy et al. for the Burgers' equation without shock and the KdV equation. For the Burgers' dataset with shock formation, we consider the dataset of Raissi et al. and for the 2D advection diffusion and the Keller-Segel equations we generate the data using standard numerical solvers in Mathematica.

In this SI we focus our attention on the performance of the algorithm to discover the correct PDE with respect to the level of (white) noise and the dataset size. We add white noise (Gaussian) on the scale of the standard deviation, $\chi$, (typically in a range between 5$\%$ and $75\%$),
\begin{equation}
    y_{noisy} = y + \chi \sigma(y) \mathcal{N}(0,1)
\end{equation}
where $\mathcal{N}$ is a Gaussian distribution. 

\subsection*{Burgers' equation}

The Burgers' equation in one dimension is given by
\begin{equation}
    \partial_t u + u u_x - \nu u_{xx} = 0, 
    \label{eq:Burgers}
\end{equation}
where $u$ is the velocity of the fluid and $\nu$ is the viscosity of the fluid. Here we consider $\nu=0.1$ and we obtain a solution without a shock wave. We use the data-set presented in Rudy et al where the equation is solved for 101 time points between 0 and 10 and 256 spatial points between -8 and 8, given a total amount of 25856 points. $N$ points are randomly sampled and $\chi\%$  white noise is added (we set the maximal number of epochs to 50000). In Fig. \ref{fig:Burgers}(a) we show, for one specific dataset of 2000 samples and 20$\%$ of noise the ground truth, noisy and reconstructed equation. In Fig. \ref{fig:Burgers}(b,c) we show for 5 randomly sampled datasets the accuracy of the algorithm as function of (left) the noise level (at $N = 500$) and (right) number of sample points (at vanishing noise levels). The number inside each square and its colour represent the mean relative error in the coefficients, if the right PDE was found. Red squares indicate the correct equation was not recovered. 

\subsection*{Korteweg-de Vries}

We consider the Korteweg-de Vries (KdV) equation as a second benchmark problem. It describes waves in shallow water and is of particularl interest for model selection due to its third order term,
\begin{equation}
    \partial_t u = 6uu_x - u_{xxx}.
    \label{Eq:KdV}
\end{equation}

We use the data-set of Rudy et al. for 201 time steps between 0 and 20 and spatial resolution of 512 cells between -30 and 30, giving a total dataset size of 120912. We randomly sample 2000 points and add $20\%$ white noise, resulting in Fig. \ref{fig:kdv}(a). This example shows that even for higher order equations the algorithm is capable of recovering the correct PDE. While for $2000$ sample points we recover the correct PDE we observe in Fig. \ref{fig:kdv}(b,c) that this can strongly depend on the sampling (we set the maximal number of epochs to $100.000$). The performance with respect to the number of sample points is slightly lower compared to the Burgers' equation due to the localized solution: the area outside the two waves is close to zero and hence does not contain information on the solution. These effects could be overcome by more advanced sampling such as latin hypercube sampling.

\subsection*{Higher dimensional input} 

To showcase the robustness of DeepMoD on high-dimensional input data, we consider the 2D advection diffusion process described by,
\begin{equation}
    \partial_t u =  -\nabla\cdot\left(-D\nabla u + \vec{v} \cdot u \right),
    \label{Eq:AdvectionDiffusion}
\end{equation}
where $\vec{v}$ is the velocity vector describing the advection and $D$ is the diffusion coefficient. Our initial condition is a 2D Gaussian and set $D = 0.5$ and $\vec{v} = (0.25,0.5)$. For as little as 5000 randomly sampled points we recover the correct form of the PDE as well as the vector $\vec{v}$ for noise levels up to $\approx 25 \%$ (See SI 2). Our library consists of all spatial derivatives (including cross terms) as well as first order polynomial terms, totalling 12 terms. At vanishing noise levels we obtain the correct equations with as little as 200 sample points through space and time. This number is surprisingly small considering this is an 2D equation. To illustrate this, Fig. \ref{fig:4} shows the ground truth, sampled points with $10\%$ noise and the reconstructed profile at two different times. Note that while it is impossible to identify the underlying process by sight, the algorithm correctly discovers the underlying PDE. After the network learned the mapping $x,t \to u$ on 5000 sample points, we can reconstruct the solution on the entire domain (Fig. \ref{fig:4} right), highlighting the ability of DeepMoD to act as a highly-accurate physics-informed interpolater and denoiser. 

\subsection*{Advection-Diffusion equation}

To model the temporal evolution of a concentration field in two-dimensions, subject to advection-diffusion, we numerically solve
\begin{equation}
    \partial_t u =  -\nabla\cdot\left(-D\nabla u + \vec{v} \cdot u \right),
\end{equation}
where $\vec{v}$ is the velocity vector describing the advection and $D$ is the diffusion coefficient. We consider a Gaussian initial condition $u(x,y)|_{t=0}= 20 \exp -\left( x^2 + y^2\right)$ over time and space where we set $D = 0.5$ and $\vec{v} = (0.25,0.5)$. For 5000 data-points randomly sampled through space (41 time-frames) we recover the correct form of the PDE as well as the vector $\vec{v}$ for noise levels up to $\approx 25 \%$ (See Fig. \ref{fig:decision_boundary_AD}). At vanishing noise levels we obtain the correct equations with as little as 200 sample points through space and time. This number is surprisingly little considering this is an 2D equation, with respect to the 1D Burgers' and KdV equations (See Fig. \ref{fig:decision_boundary_AD}). 

\section*{Gel electrophoresis}

\subsection{Advection-Diffusion}

To test the performance of DeepMoD on experimental data we perform a gel electrophoresis experiment with a (purple) charged loading dye ($10 \mu l$), inserted inside an agarose gel (1g agarose /100ml TAE buffer) and tracked in an electrophoresis assay. A low voltage of 12 V is applied (lowest voltage the power-supply allows). For this voltage, the electric field drives the charged loading dye with a constant velocity through the gel. For this low voltage, the advective motion due to the electric field is accompanied by a diffusive motion of the dye in the gel which can both be observed in Fig. 5 of the main text.

We analyse the density profile of two initially localised dyes (See Fig. 5 in the main text) to infer the underlying advection-diffusion equation (Eq. \ref{Eq:AdvectionDiffusion}, using a library of 12 candidate terms, similarly as for the artificial advection diffusion data. The  neural network we used contained 3 hidden layers of 20 neurons ($\lambda = 10^{-5}$). 5000 data-points are randomly sampled through space ($x= 300$ pixels in the direction of advection and  $y= 150$ pixels perpendicular) and time (50 time-frames with 10 min interval). As discussed in the main text, an automatic threshold of the coefficient vector does not render a sparse solution due to small yet non-zero higher order terms. Therefore we fix the threshold to $|\xi| < 0.1$. In  Table. S1 we present the resultant equation for both experiments with five different sampling seeds. In 8/10 cases the correct equation is obtained, while in 2 cases two small, yet non-zero additional $u u_x$, $u u_y$ terms are found. To illustrate the robustness of the algorithm with respect to the precise architecture we show in Table SI2 that increasing or decreasing the number of layers in the neural network did not yield different results (except for one single layer, which is not sufficient to serve as function approximator). 

\begin{table}
\begin{tabular}{ |p{2cm}||p{7.15cm}|p{7.15cm}|  }
 \hline
 \multicolumn{3}{|c|}{Electrophoresis experiment} \\
 \hline
Run & Experiment 1 (solid line) & Experiment 2 (dashed line)\\
 \hline
1   & $u_{t} = 0.31 u_y + 0.011 u_{xx} + 0.009 u_{yy}$    &$u_{t} = 0.28 u_y + 0.007 u_{xx} + 0.016 u_{yy} + 0.001 u u_{xx} - 0.002 u u_{yy}$   \\
2&   $u_{t} = 0.30 u_y + 0.009 u_{xx} + 0.009 u_{yy}$  & $u_{t} = 0.28 u_y + 0.009 u_{xx} + 0.009 u_{yy}$  \\
3 &$u_{t} = 0.31 u_y + 0.010 u_{xx} + 0.010 u_{yy}$ & $u_{t} = 0.29 u_y + 0.008 u_{xx} + 0.011 u_{yy}$\\
4   &$u_{t} = 0.31 u_y + 0.010 u_{xx} + 0.011 u_{yy}$  & $u_{t} = 0.29 u_y + 0.009 u_{xx} + 0.011 u_{yy}$\\
5&   $u_{t} = 0.31 u_y + 0.007 u_{xx} + 0.007 u_{yy}$   & $u_{t} = 0.29 u_y + 0.008 u_{xx} + 0.015 u_{yy}+ 0.001 u u_{xx} - 0.002 u u_{yy}$ \\

 \hline
\end{tabular}
\caption{Equations obtained for the two electrophoresis experiment (See Fig. 5 in the main text) for five different seeds. The red terms that are indicated are the terms that in 2/10 cases are incorrectly retained after treshholding the coefficient vector.}
\label{table:1}
\end{table}

\begin{table}
\centering
\begin{tabular}{|p{4.15cm}|p{7.15cm}|  }
 \hline
 \multicolumn{2}{|c|}{Impact of neural network architecture} \\
 \hline
Layers & Result \\
 \hline
 1  & $u_t =  0.31 u_y +0.19 - 0.16 u$   \\
2  & $u_{t} = 0.30 u_y + 0.009 u_{xx} + 0.008 u_{yy}$  \\
3 & $u_{t} = 0.30 u_y + 0.009 u_{xx} + 0.009 u_{yy}$\\
4 & $u_{t} = 0.30 u_y + 0.009 u_{xx} + 0.009 u_{yy}$\\
5 & $u_{t} = 0.30 u_y + 0.009 u_{xx} + 0.009 u_{yy}$ \\

 \hline
\end{tabular}
\caption{Equations obtained for the two electrophoresis experiment (See Fig. 5 in the main text) for networks with different amount of layers consisting of 20 neurons with a tanh activation function. The red terms are incorrect. Note that for at least two layers, the size of the network has neglegible effect on the results.}
\label{table:2}
\end{table}

\begin{figure}
    \centering
    \includegraphics[width=0.5 \linewidth]{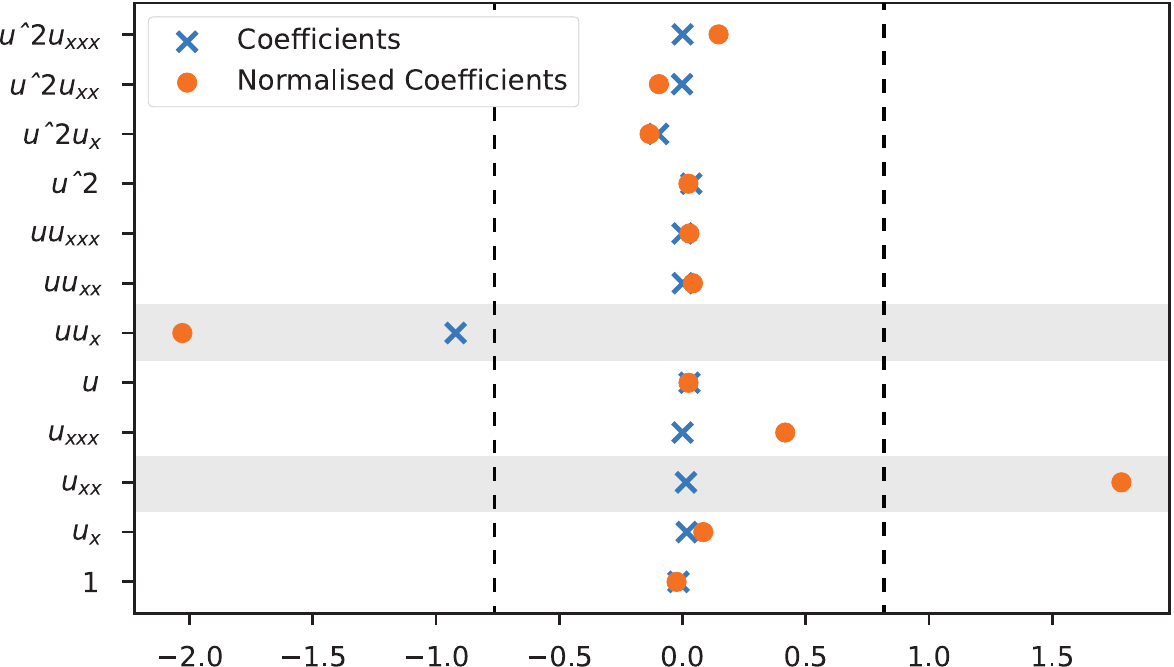}
    \caption{Scaled (orange dots) and unscaled (blue crosses) components of $\xi$ after $10^5$ epochs of training on the Burgers with shock dataset of 2000 samples and $10\%$ white noise. The grey bars denote the terms to be discovered and the vertical dashed lines the area which will be thresholded.}
    \label{fig:scaling}
\end{figure}

\begin{figure}
    \centering
    \includegraphics[width=0.65 \linewidth]{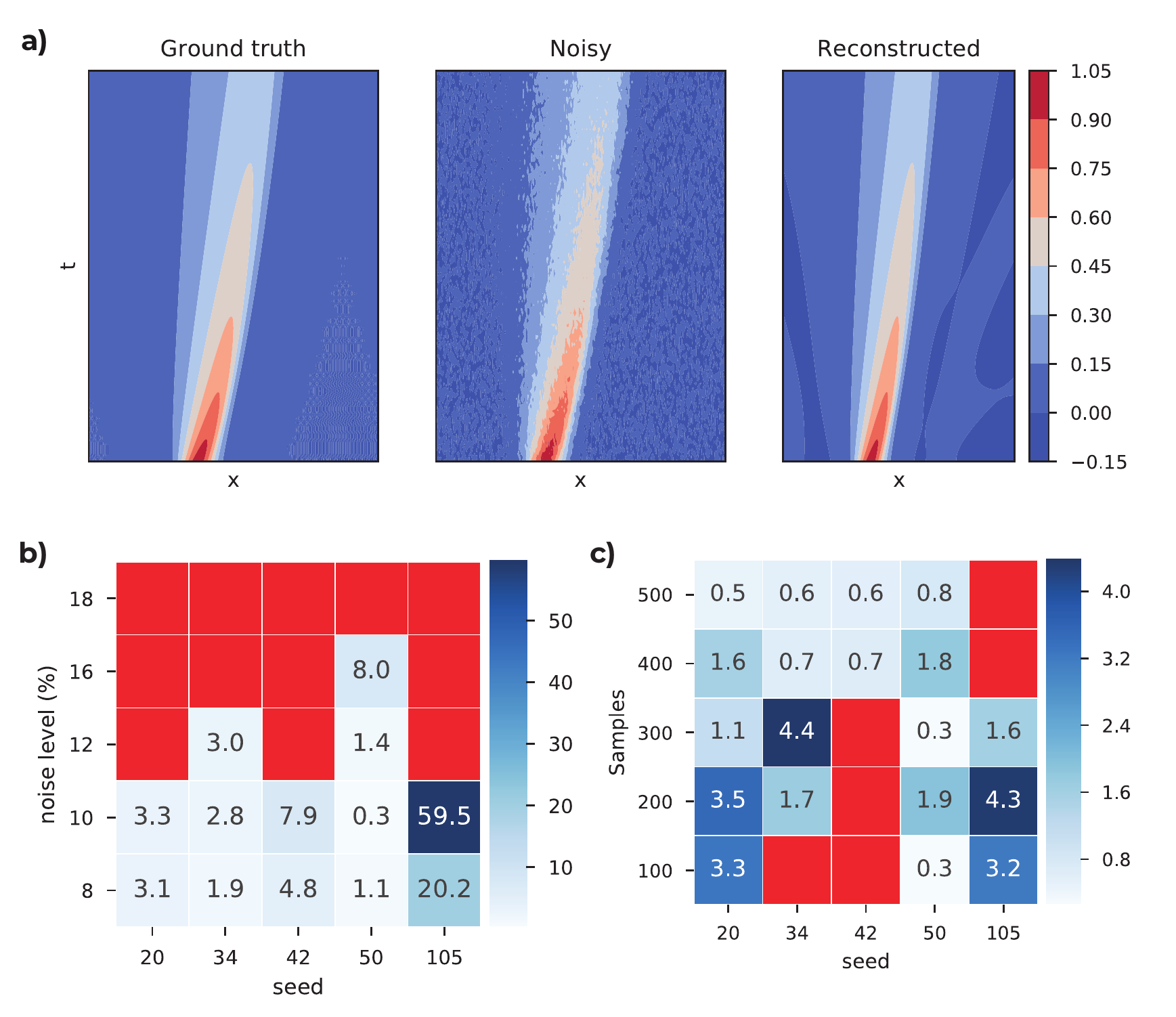}
    \caption{a) Ground truth (left), Noisy (middle) and reconstructed (right) solution for the Burgers' equation. b) Accuracy of the algorithm for the Burgers' equation as function of the noise levels for $500$ sample points and c) as function of sample size (vanishing noise levels). Annotation and colors indicate the mean error on the coefficients, if the correct PDE was found. Red squares indicate the right PDE was not found.}
    \label{fig:Burgers}
\end{figure}

\begin{figure}
    \centering
    \includegraphics[width=0.65 \linewidth]{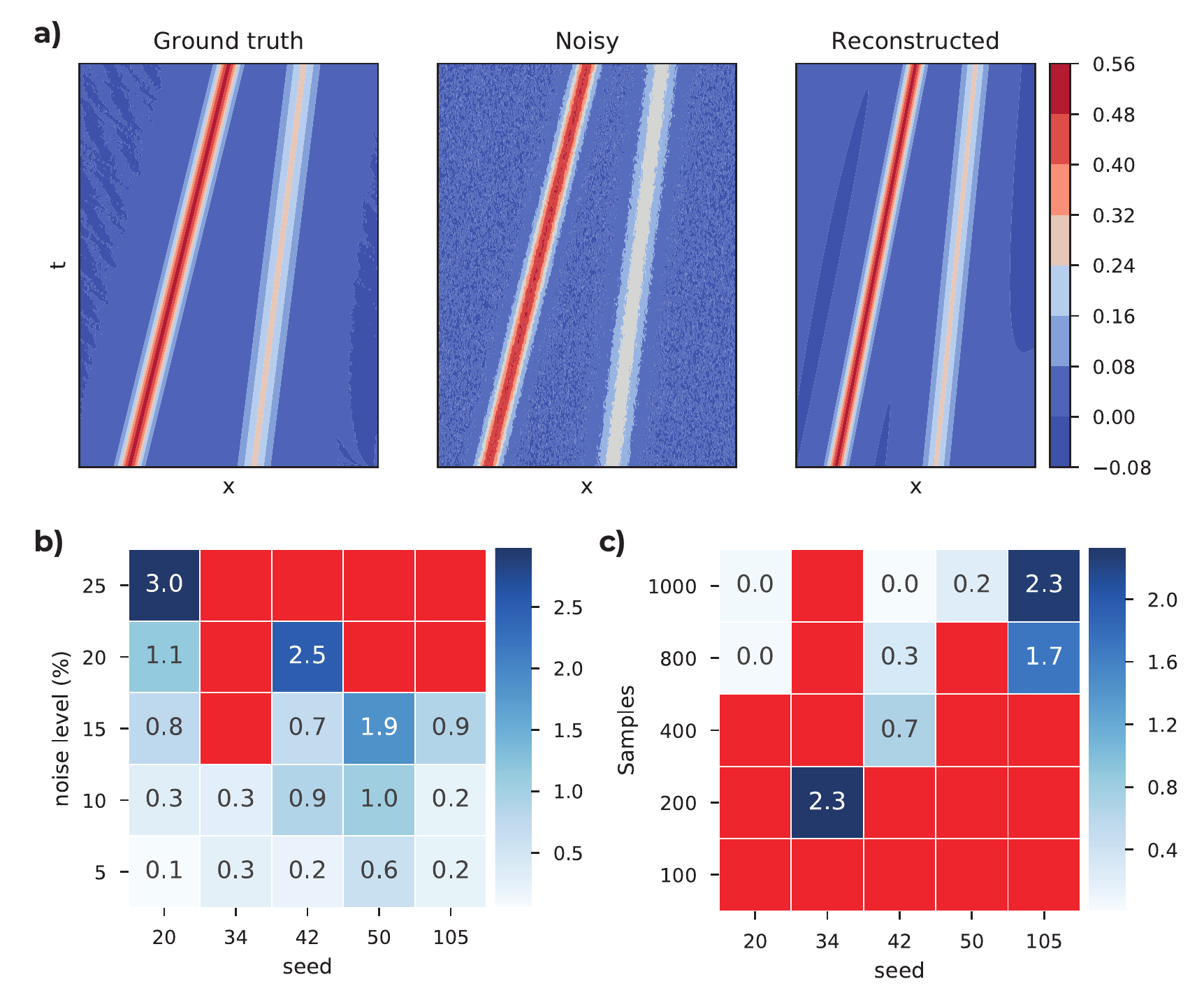}
    \caption{a) Ground truth (left), Noisy (middle) and reconstructed (right) solution for the Korteweg-de Vries equation. Accuracy of the algorithm for the Korteweg - de Vries equation (b) as function of of the noise levels for $2000$ sample points and (c) as function of sample size (vanishing noise levels). Annotation and colors indicate the mean error on the coefficients, if the correct PDE was found. Red squares indicate the right PDE was not found.}
    \label{fig:kdv}
\end{figure}

\begin{figure}
    \centering
    \includegraphics[width=0.65 \linewidth]{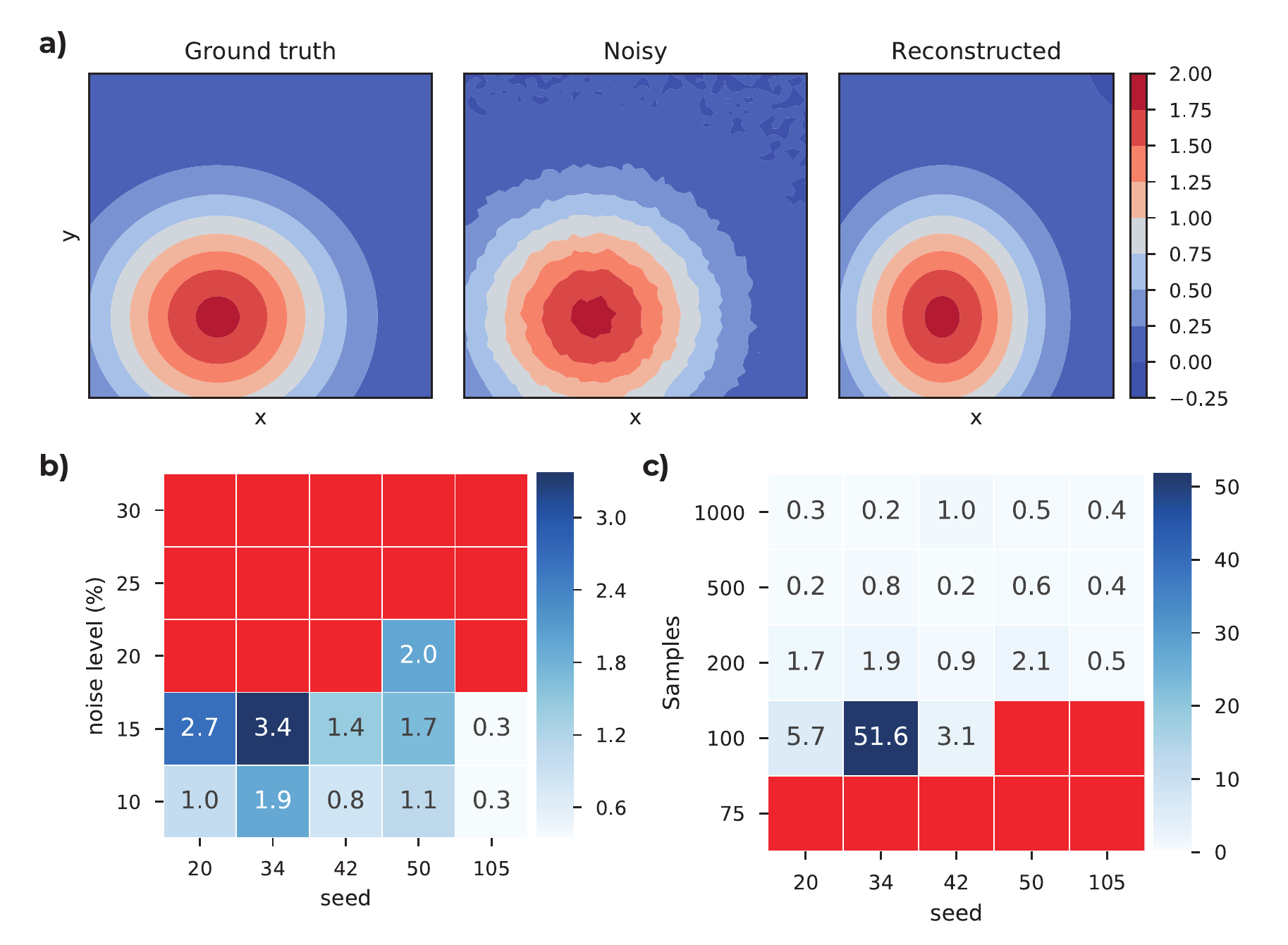}
    \caption{a) Ground truth (left), Noisy (middle) and reconstructed (right) solution for the 2D advection diffusion equation. Accuracy of the algorithm for the 2D advection diffusion equation, (b) as function of of the noise levels for $5000$ sample points and (c) as function of sample size (vanishing noise levels). Annotation and colors indicate the mean error on the coefficients, if the correct PDE was found. Red squares indicate the right PDE was not found.}
    \label{fig:decision_boundary_AD}
\end{figure}

\begin{figure}
\centering
\includegraphics[width=0.8 \linewidth]{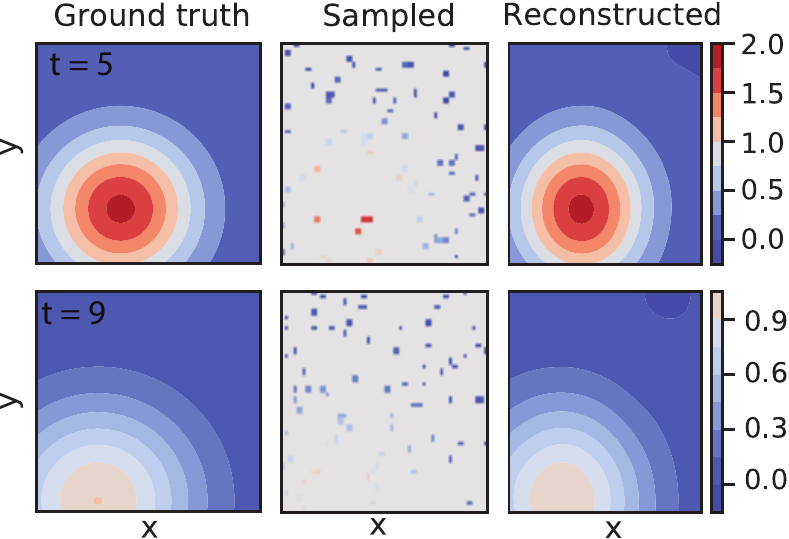}
\caption{Ground truth (left), Sampled (middle) and reconstructed (right) solution for the 2D advection diffusion equation at $t= 5,9$ (10 $\%$ white noise and 5000 samples). Note that, even for an extremely low sampling rate, the full concentration profile is recovered, as well as the corresponding PDE of the process.}
\label{fig:4}
\end{figure}